\documentclass[apl,twocolumn,amsmath,amssymb,superscriptaddress,floatfix]{revtex4}

\usepackage{graphicx}
\usepackage{dcolumn}
\usepackage{bm}
\usepackage[sort&compress]{natbib}
\usepackage{xspace}
\usepackage{amsmath}
\usepackage{amssymb}
\usepackage{epsfig}

\newcommand{\mum}{ \,\mu\text{m}}

\begin{document}
\title[Davan\c co \it{et al.},"Efficient quantum dot single photon extraction into an optical fiber..."]{Efficient quantum dot single photon extraction into an optical fiber using a nanophotonic directional coupler}

\author{M. Davan\c co} \email{mdavanco@nist.gov}
\affiliation{Center for Nanoscale Science and Technology, National
Institute of Standards and Technology, Gaithersburg, MD 20899, USA}
\affiliation{Maryland NanoCenter, University of Maryland, College
Park, MD 20742, USA}
\author{M. T. Rakher}
\affiliation{Center for Nanoscale Science and Technology, National
Institute of Standards and Technology, Gaithersburg, MD 20899, USA}
\author{W. Wegscheider}
\author{D. Schuh}
\affiliation{Institute for Experimental and Applied Physics,
University of Regensburg, D-93053 Regensburg, Germany}
\author{A. Badolato}
\affiliation{Department of Physics and Astronomy, University of
Rochester, Rochester, New York 14627, USA}
\author{K. Srinivasan}
\affiliation{Center for Nanoscale Science and Technology, National
Institute of Standards and Technology, Gaithersburg, MD 20899, USA}

\begin{abstract} We demonstrate a spectrally broadband and efficient
technique for collecting emission from a single InAs quantum dot
directly into a standard single mode optical fiber.  In this
approach, an optical fiber taper waveguide is placed in contact with
a suspended GaAs nanophotonic waveguide with embedded quantum dots,
forming a broadband directional coupler with standard optical fiber
input and output. Efficient photoluminescence collection over a
wavelength range of tens of nanometers is demonstrated, and a
maximum collection efficiency of 6~\% (corresponding single photon
rate of 3.0 MHz) into a single mode optical fiber is estimated for a
single quantum dot exciton.
\end{abstract}

\pacs{78.55.-m, 78.67.Hc, 42.70.Qs, 42.60.Da}

\maketitle

Single epitaxially grown quantum dots (QDs) can serve as bright,
stable sources of single photons for applications in quantum
information processing~\cite{ref:Shields_NPhot}. A key limitation of
QDs embedded in high refractive index semiconductors is the
relatively small fraction of the total QD emission
($\approx1~\%$)~\cite{ref:Benisty3,ref:Barnes2} that can be
collected in free-space by a high numerical aperture optic, a
consequence of total internal reflection at the semiconductor-air
interface. Embedding the QD in a high quality factor ($Q$), small
mode volume resonator such as a micropillar
cavity~\cite{ref:Solomon,ref:Pelton} is one approach to improving
photon extraction, where ideally one benefits from both a faster
radiative rate (Purcell enhancement) and a far-field emission
pattern that can be efficiently collected. High extraction
efficiencies have indeed been demonstrated with this
approach~\cite{ref:Strauf_NPhot}. One challenging aspect of using a
high-$Q$ microcavity is the necessity for spectral overlap between a
narrow cavity mode and the QD emission line, though tunable
geometries~\cite{ref:Muller_APL} can overcome this challenge.
Alternatively, spectrally broadband approaches (usually without
Purcell enhancement) avoid precise tuning and are needed for
efficient spectroscopy of multiple spectrally distinct QD
transitions and/or emission from multiple QDs, and have recently
been pursued using solid immersion
lenses~\cite{ref:Zwiller_Bjork_JAP,ref:Vamivakas_NL} and in
vertically oriented tapered nanowire geometries~\cite{ref:Claudon}.
Here, we demonstrate a guided wave nanophotonic structure for
efficient extraction of PL from a single InAs QD directly into an
optical fiber, with an operation bandwidth of tens of nm, and an
overall single mode fiber collection efficiency of $\approx$6~\%.
Since collection is directly into an optical fiber, optical losses
associated with light extraction using a typical
micro-photoluminescence setup and coupling from free-space into an
optical fiber are completely avoided. Furthermore, this geometry is
planar, avoiding deeply etched, vertically-oriented geometries, and
could serve as a platform for waveguide-based photonic circuits
involving single QDs and efficient coupling to optical fibers.

\begin{figure}
\centerline{\includegraphics[width=0.97\linewidth,trim= 0 10 0
10]{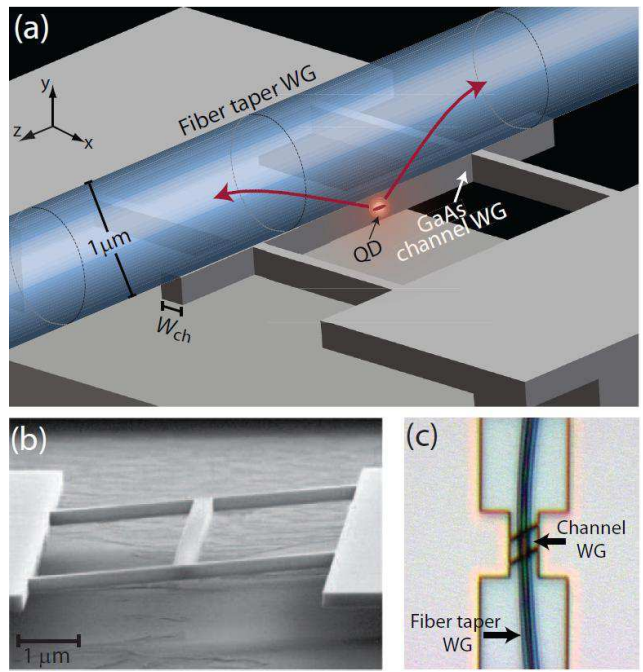}} \caption{(a) Nanophotonic directional coupler for
single photon extraction from a single embedded QD. (b) SEM image of
a fabricated GaAs channel WG. (c) Optical microscope image of the
FTW/channel WG directional coupler.} \label{Fig:RTprobing}
\end{figure}

\begin{figure*}
\begin{center}
\begin{minipage}[c]{0.68\linewidth}
\includegraphics[width=\linewidth]{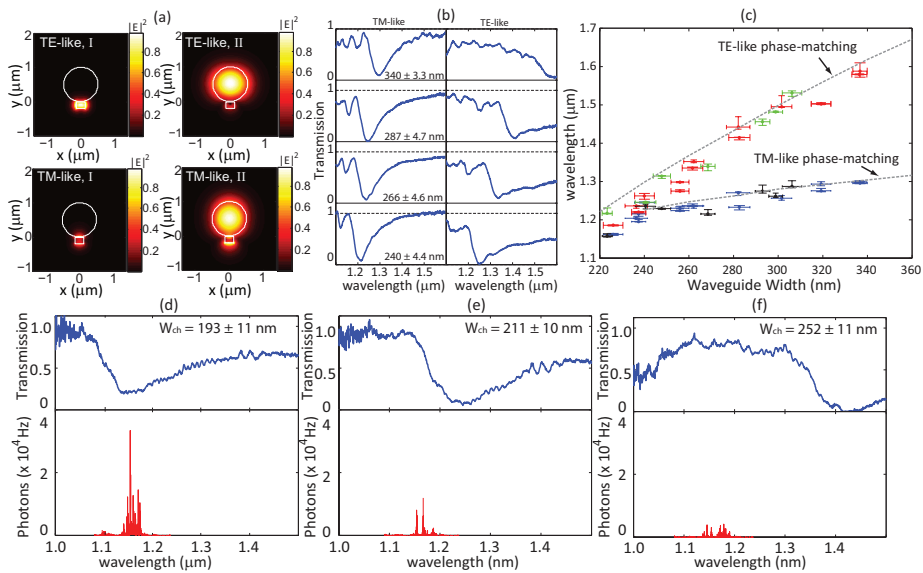}
\end{minipage}\hfill
\begin{minipage}[c]{0.29\linewidth}
\caption{(a) Electric field amplitude squared for the TE-like (top)
and TM-like (bottom) supermode pairs of the directional coupler. (b)
TE-like and TM-like transmission spectra for suspended channel WGs
of varying widths, probed with a $\approx1~\mu$m FTW at room
temperature. (c) Evolution of transmission minima with WG widths for
the two polarizations in (a). Dashed lines are calculated
phase-matching wavelengths for the fiber and WG. (d)-(f): Low
temperature ($\approx$8 K) transmission and corresponding
fiber-collected PL for WGs with increasing widths $W_{ch}$.}
\label{Fig:WLspex}
\end{minipage}
\end{center}
\end{figure*}

Our structure is a hybrid directional coupler formed by a suspended
GaAs channel waveguide (WG) containing InAs QDs and a micron
diameter optical fiber taper waveguide  (FTW)
(Fig.~\ref{Fig:RTprobing}(a)). The FTW is an optical fiber whose
diameter is adiabatically reduced to a wavelength scale minimum,
providing access to an evanescent field for guided wave coupling
while maintaining single mode fiber ends and low loss. The GaAs WG
(Fig.~\ref{Fig:RTprobing}(b)) has a cross-sectional diameter of
$\approx100$ nm, enabling phase matching to the FTW and efficient
power transfer between the two WGs~\cite{ref:Davanco2}. This
structure supports single guided modes with strong transverse
confinement, into which QD radiation is almost completely coupled.
Efficient QD coupling to WG modes phase-matched to the single FTW
mode thus leads to efficient extraction of QD emission into the
fiber. Detailed simulations have predicted a single QD fluorescence
collection efficiency as high as $\approx35~\%$ into an optical
fiber ($\approx 70~\%$ including both fiber ends), with an operation
bandwidth of tens of nm~\cite{ref:Davanco2}.

We first assessed the directional coupler without QDs, to confirm
the basic light transfer mechanism between FTW and channel WG. A
first set of suspended WGs with no QDs was fabricated on a 250~nm
thick GaAs wafer for passive directional coupler
characterization~\cite{wg_apl_note}. An $\approx 1~\mu$m diameter
FTW was brought into contact with individual channel WGs, forming
directional couplers as illustrated in Fig.~\ref{Fig:RTprobing}(c).
Transmission spectra were obtained by launching broadband polarized
light from a tungsten halogen lamp into the FTW input and analyzing
output light with an optical spectrum analyzer.  The FTW and channel
WG each support a single guided mode of TE-like (x-oriented electric
field) and TM-like (y-oriented electric field) polarizations.  The
resulting directional coupler supports a pair of hybrid supermodes
for each polarization (Fig.
\ref{Fig:WLspex}(a))~\cite{ref:Davanco2}. The transmission spectrum
for a given polarization is determined by the beating of the
corresponding coupler supermodes, and exhibits minima when power is
transferred from the FTW to the channel WG but not back into the FTW
due to termination of the channel~\cite{wg_apl_note}.

Several 8 $\mu$m long WGs with widths between $240$~nm and $340$~nm
were measured. The transmission spectra (Fig.~\ref{Fig:WLspex}(b))
for the two main coupler polarizations displayed broad, $>40$~nm
wide minima which typically reached $>90~\%$ extinction, evidencing
efficient power transfer between the FTW and suspended WGs. After
optical characterization, WG widths were measured with a scanning
electron microscope. Figure~\ref{Fig:WLspex}(c) shows minimum
transmission wavelengths as a function of WG width, along with the
phase-matching wavelengths calculated with a vector finite element
method~\cite{ref:Davanco2}. The minima closely follow the calculated
phase-matching wavelengths. The higher rate with which the TE-like
phase-matching wavelength shifts with WG width is expected from
these modes' higher intensity at the WG sidewalls. The agreement
between theoretical and experimental curves indicates that the
expected efficient directional coupler operation is indeed achieved.

We next attempted to validate efficient PL extraction from a second
set of devices fabricated on a high QD density portion of the same
wafer, where the QD ensemble s-shell emission peak was at
$\approx$1200~nm. The QD-containing WGs were probed with a $\approx
1~\mu$m diameter FTW in a cryostat at $< 9$~K temperature with the
setup shown in Fig. S-3 of the supporting
materials~\cite{wg_apl_note}. The devices were excited by launching
a 50~MHz repetition rate, 50~ps width, 780~nm laser pulse train into
the FTW. Figures~ \ref{Fig:WLspex}(d)-(f) show TE polarization
transmission and corresponding PL spectra for three devices of
varying WG widths. PL collection over a range of a few tens of nm is
achieved in all cases. The collected PL is maximized when the
transmission dip - and thus the phase-matching wavelength - is
aligned with the s-shell PL peak, evidencing efficient power
transfer between phase-matched WGs. Fiber-collected PL of individual
emission lines was typically 10 to 100 times higher than that
obtained via free-space collection (from the same devices) with a
0.42 numerical aperture objective. Estimating the absolute
collection efficiency was difficult, however, as the high density of
QD lines prevented accurate determination of the intensity of any
individual transition.

\begin{figure*}[!]
\begin{center}
\begin{minipage}[c]{0.68\linewidth}
\includegraphics[width=\linewidth]{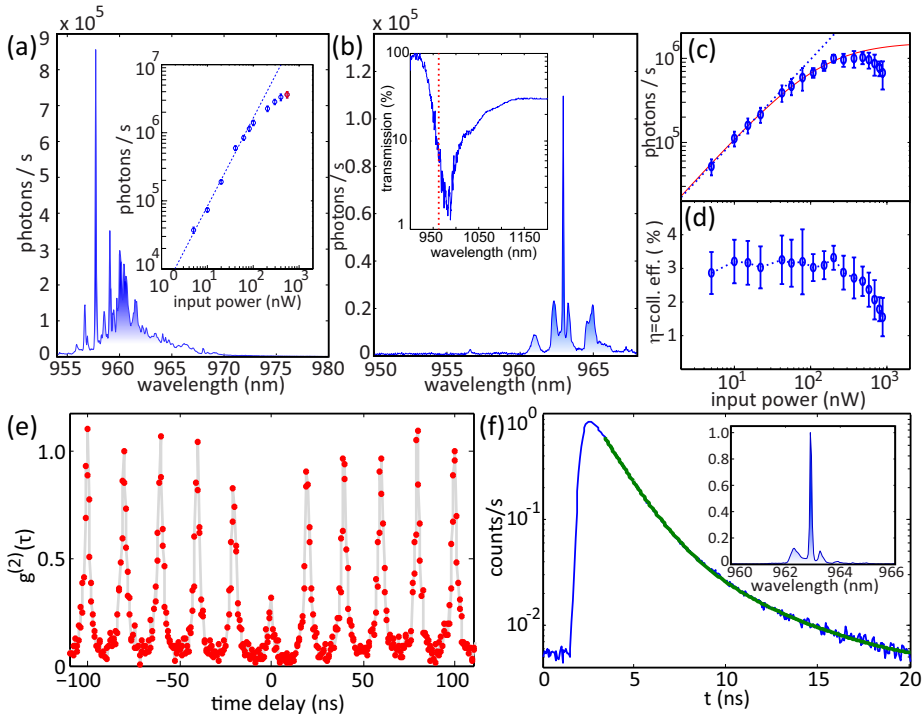}
\end{minipage}\hfill
\begin{minipage}[c]{0.29\linewidth}
\caption{(a)Fiber-collected single QD PL for the brightest device.
Inset: photon rate for 957.7 nm line against pump power. (b)
Fiber-collected single QD PL for second WG. Inset: directional
coupler transmission spectrum, showing position of QD line. (c)
Collected PL rate versus excitation power for 963 nm line in (b).
Error bars are 95~\% fit confidence intervals. Dashed line: linear
fit to data below saturation. Continuous line: fit to theory. (d)
Fiber collection efficiency obtained from (c), assuming ideal QD
behavior. (e) Second-order correlation $g^{(2)}(\tau)$ and (f)
lifetime for the excitonic line in (b), after a 1~nm bandpass filter
(inset spectrum). Green line: bi-exponential fit.} \label{Fig:SQDPL}
\end{minipage}
\end{center}
\end{figure*}

To avoid this difficulty, a third set of devices was fabricated with
a low density of QDs, so that well-isolated transitions could be
observed. These devices were produced from a 190~nm thick, GaAs WG
layer wafer containing QDs with ensemble s-shell emission near
940~nm. Low-temperature PL spectroscopy was performed as above, with
a 50~MHz pulsed pump at 780~nm. Figure~\ref{Fig:SQDPL}(a) shows the
PL spectrum from the device that yielded the highest collection
efficiency, for the sharp excitonic line at 957.7~nm. Driving this
transition towards saturation (inset, Fig.~\ref{Fig:SQDPL}(a)), at
550~nW pump power we estimate a collection efficiency into a single
mode optical fiber $\eta=6.05~\%\pm0.061~\%$~\cite{wg_apl_note}. We
point out that the transmission of the collection portion of the FTW
is $\approx84~\%$, so collection into the FTW (our first collection
optic) is $7.250~\%\pm0.072~\%$. The most likely reason for smaller
collection efficiencies than predicted is non-optimal positioning of
the QD in the GaAs WG, as supported by simulations presented in the
supplementary material~\cite{wg_apl_note}.

The 957.7~nm transition was located outside the operating wavelength
band of the tunable bandpass filter available for spectral
isolation, preventing further characterization of this device. A
second WG was available, however, that displayed an isolated
transition at $\approx963$~nm, shown in Fig.~\ref{Fig:SQDPL}(b),
located within the coupler transmission dip (inset). The evolution
of the 963~nm peak collected PL rate into the single mode fiber as a
function of pump power is shown in Fig.~\ref{Fig:SQDPL}(c). For
$P_{in}<100$~nW, integrated PL counts increase linearly with pump
power. A fit to the data assuming an ideal single exciton QD line
(Fig.~\ref{Fig:SQDPL}(c))~\cite{wg_apl_note} matches the data for
$P_{in}<100$~nW, but overestimates the PL rate for $P_{in}>100$~nW.
For the highest measured PL intensity, the integrated counts
correspond to $\eta\approx2~\%$. For $P_{in}\ll
P_{sat}\approx133$~nW, where the QD can be assumed to behave
ideally, however, $\eta\approx3~\%$ (Fig. \ref{Fig:SQDPL}(d)).

To confirm the single photon nature of the 963~nm line of
Fig.~\ref{Fig:SQDPL}(b), we measured the second-order correlation,
$g^{(2)}(\tau)$, for pulsed excitation at $P_{in}\approx 75$~nW
(Fig.~\ref{Fig:SQDPL}(e)).
$g^{(2)}(0)=0.29\pm0.02$~\cite{wg_apl_note}, indicating single QD
emission that is dominantly comprised of single photons. The nonzero
$g^{(2)}(0)$ is likely due to insufficient filtering, which allows
uncorrelated photons and possible emission from other QDs to be
detected. In particular, the spectrum of the detected light (inset
in Fig.~\ref{Fig:SQDPL}(f)) contains two broad sidelobes in addition
to the 963 nm excitonic line. These may correspond to emission from
other QDs with broadened lines, owing to proximity to WG
sidewalls~\cite{wang_APL_3423}. A bi-exponential decay of the
excitonic line (Fig.~\ref{Fig:SQDPL}(f)) with fast and slow
lifetimes of $1.48$~ns$\pm0.08$~ns and $4.6$~ns$\pm0.8$~ns
(uncertainties are 95~\% fit confidence intervals) evidences
non-ideal QD behavior. This is consistent with what is seen in
$g^{(2)}(\tau)$ (Fig.~\ref{Fig:SQDPL}(e)), where the coincidence
counts between peaks do not return all the way to zero.  The fast
decay constant approaches the lifetime of a typical InAs QD (the
lack of radiative rate enhancement is predicted in
simulations~\cite{ref:Davanco2}), while the long decay evidences QD
coupling to nonradiative states that may lead to a reduced quantum
efficiency and collection efficiency estimates.

In summary, we have demonstrated a fiber-coupled, QD single photon
source based on a planar, guided wave nanophotonic coupler.  We use
this spectrally broadband approach to demonstrate an in-fiber,
single QD PL collection efficiency of 6~\%.  Future work is aimed at
improved efficiency through precise QD
location~\cite{ref:Hennessy3,ref:Thon} within the device and
understanding sources of non-ideal QD behavior.

\acknowledgements This work was partly supported by the
NIST-CNST/UMD-NanoCenter Cooperative Agreement.  We thank R. Hoyt,
C. S. Hellberg, Alexandre M.P.A. Silva, and Hugo
Hern\'andez-Figueroa for useful discussions.

\bibliographystyle{apsrev}


\onecolumngrid \bigskip
\appendix
\setcounter{figure}{0}
\begin{center} {{\bf \large SUPPORTING
INFORMATION}}\end{center}

\section*{{Finite difference time domain simulations}} Finite
difference time domain simulations were used to study the
relationship between coupler transmission and PL spectra. The
simulated structure consisted of a 1~$\mum$ diameter FTW of infinite
extent and a 190~nm thick, 160~nm wide, 4~$\mum$ long GaAs channel
waveguide. The computational domain was terminated with perfectly
matched layers to simulate open boundaries.

\subsection*{Transmission spectrum}
To determine the simulated transmission spectrum, a fundamental mode
was launched into the FTW, at a 3~$\mum$ distance from the channel
waveguide, and steady-state fields after the channel waveguide
termination (for transmission) and before the mode launch position
(for reflection), were recorded at various wavelengths. These fields
were then convolved with the FTW mode and normalized to the injected
power, to yield transmission and reflection spectra.
Fig.~\ref{Fig:T_and_R}(a) shows transmission and reflection spectra
for the structure with the above dimensions, and (b)-(d) show
electric field profiles, recorded at the $x=0$ plane, for FTW mode
excitation at $z=-3~\mum$ at three different wavelengths. A broad
transmission dip, and corresponding reflection peak, are observed in
Fig.~\ref{Fig:T_and_R}(a) at $\lambda=1072$~nm, where FTW and
channel waveguide are phase-matched. In the corresponding field
profile, Fig.~\ref{Fig:T_and_R}(c), it is evident that efficient
power transfer occurs from the FTW into the waveguide, along the
4~$\mum$ extent of the latter. Due to waveguide termination at
$z=2~\mum$, however, transfer back into the optical fiber is
interrupted. Power guided in the channel is, instead, strongly
scattered at the waveguide facet, and, as evidenced by the
standing-wave pattern at $z<0$, partially reflected back into the
fiber. At wavelengths away from the phase-matching condition,
Fig.~\ref{Fig:T_and_R} (b) and (d), power launched into the fiber is
transferred, without significant losses, past the GaAs guide, and
transmission tends to $100$~\%, reflection to 0.  Comparing the
calculated transmission spectrum of Fig.~\ref{Fig:T_and_R} to the
measured curves in Fig.~2 in the main text, we see that the measured
data matches prediction well in terms of their general shape and in
particular, their depth and bandwidth.

\subsection*{PL spectrum}

\begin{figure}[t]
\centerline{\includegraphics[width=15cm]{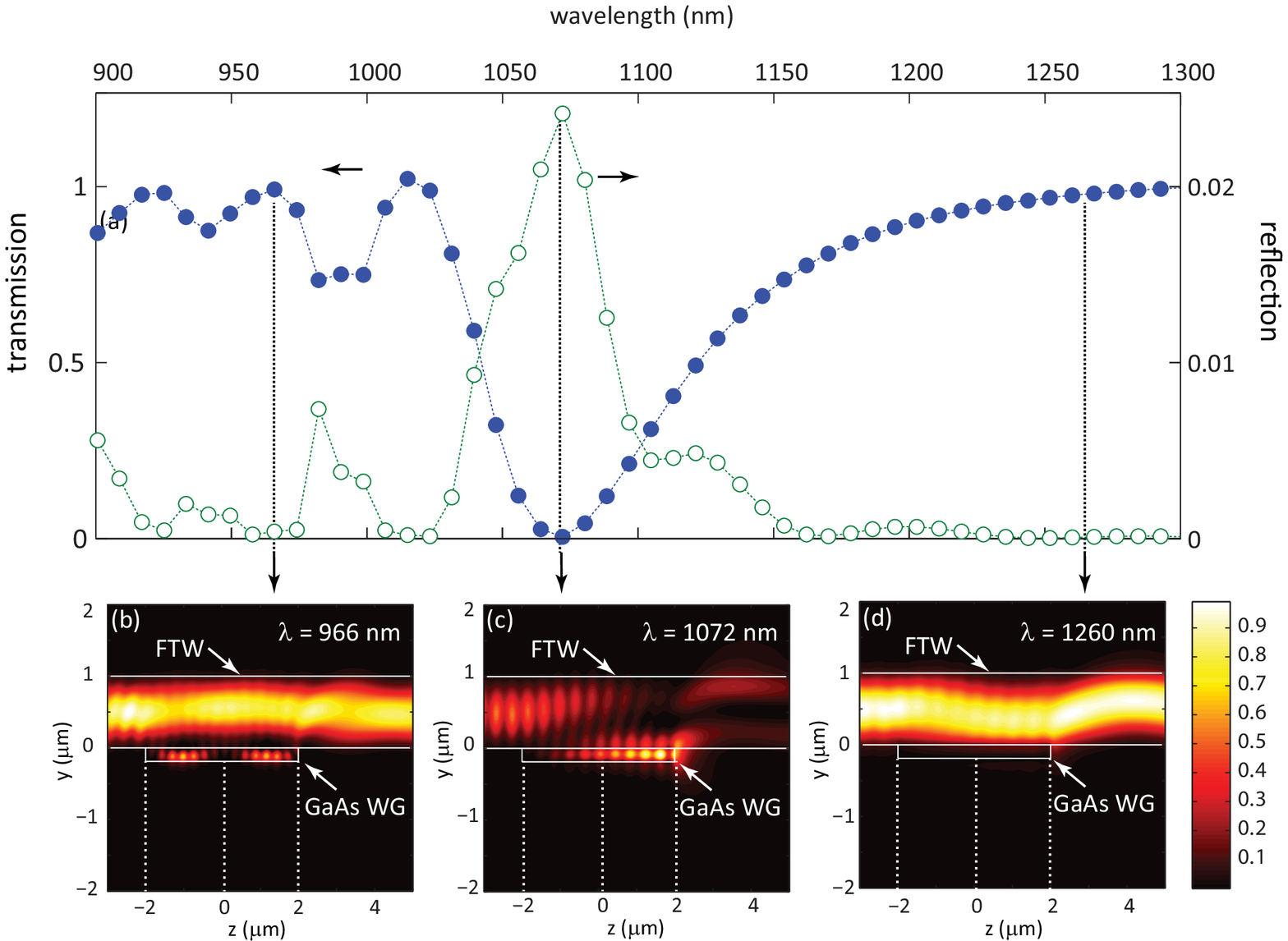}}
\caption{(a) Transmission (blue filled circles) and reflection
(green open circles) spectra for simulated directional coupler
formed by 1~$\mum$ diameter FTW and a 190~nm thick, 160~nm wide,
4~$\mum$ long suspended GaAs WG. (b)-(d) Steady-state, amplitude
squared electric field at the $x=0$ plane for fiber mode excitation
at $z=-3~\mum$ at three different wavelengths. The coupler extends
along the $\hat{z}$ direction, and the 4$~\mum$ long suspended GaAs
WG is centered at $z=0$.} \label{Fig:T_and_R}
\end{figure}

\begin{figure}[t]
\centerline{\includegraphics[width=15cm]{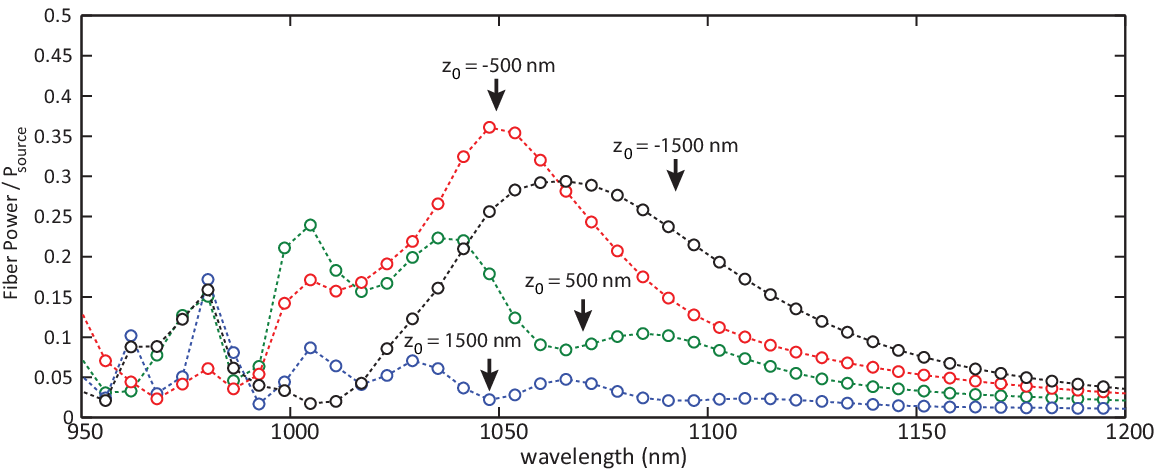}}
\caption{Predicted single QD photoluminescence collection efficiency
into the optical fiber mode for the coupler from
Fig.~\ref{Fig:T_and_R}. The curves were obtained by simulating a
horizontally oriented electric dipole at the center of the channel
waveguide cross-section ($x=0$,$y=0$), and at various positions
$z_0$ along the longitudinal direction, $\hat{z}$ ($z = 0$ is at the
center of the coupler, as in Fig.~\ref{Fig:T_and_R}).
 } \label{Fig:PL_sim}
\end{figure}

To determine simulated PL spectra, the structure was excited with a
broadband electric dipole source, and the steady-state
electromagnetic field at $z=2~\mum$ was recorded for various
wavelengths. These fields were convolved with the isolated FTW mode
field as in ref.~\onlinecite{ref:Davanco} to yield the fiber-coupled
power. The total radiated power was also recorded, and was used to
obtain the total collection efficiency. Figure~\ref{Fig:PL_sim}
shows collection efficiency for dipoles located at varying positions
$z_0$ along the coupler length, with $z=0$ at the center. Maximized
collection collection is observed at wavelengths near 1050 nm,
corresponding to the transmission dip in Fig.~\ref{Fig:T_and_R}, for
$z_0=-500$ nm. While efficient power transfer from the WG to the
fiber occurs due to phase-matching, the coupler length must be long
enough for sufficient transfer to occur, as evidenced by the
considerably lower collection efficiencies obtained for the
additional dipole positions simulated. For example, at $z_0=1500$
nm, the efficiency dips considerably below 10~\% within the coupler
operation band at $\lambda\approx1050$ nm.  The sensitivity of the
extraction efficiency with dipole position along the coupler length
is not surprising, given that it is in complete accordance with how
a waveguide directional coupler is expected to operate.  It does,
however, represent a potential reason why theoretically-predicted
efficiencies have not been achieved in experiment.

PL extraction efficiency as a function of QD dipole orientation was
studied numerically in ref.~\onlinecite{ref:Davanco_2}.  The optimal
dipole orientation is along the dominant polarization direction of
the TE-like coupler supermodes, which is the $\hat{x}$-axis as
defined in Fig.~1(a) in the main text.  There, it was shown that if
the dipole alignment was aligned along the $\hat{z}$-axis, the
predicted PL extraction dropped by about a factor of 7.  Thus, along
with the QD's position, its dipole orientation may also reduce the
measured collection efficiencies relative to the maximum possible
values.

\section*{Experimental Details}
\subsection*{Fabrication}
Wafers were grown by molecular beam epitaxy, with an epistructure
consisting of a GaAs waveguide layer on top of a 1 $\mu$m thick,
Al$_x$Ga$_{1-x}$As ($x>0.7$) sacrificial layer.  Suspended GaAs WGs
(Fig.~1(b) in the main text) containing self-assembled InAs QDs were
fabricated using standard, submicron III-V processing techniques.
Two different wafers were used, with waveguide layer thicknesses of
250 nm and 190 nm, and in the center of the waveguide a single layer
of InAs QDs with a variable density gradient (from
$>100~\mu$m$^{-2}$ to 0$~\mu$m$^{-2}$, along the $(01\bar{1})$ wafer
direction) was grown, allowing for the creation of devices with
varying QD densities.

Device fabrication was as follows. Electron-beam lithography with
ZEP520-A resist was used to define waveguide patterns on top of a
200 nm SiN$_x$ layer deposited on the epiwafer via plasma-enhanced
chemical vapor deposition (PECVD). The patterns were transferred to
the SiN$_x$ through reactive ion etching (RIE) with a CHF$_3$/Ar
mixture. Following resist removal, the waveguide patterns were
transferred to GaAs through inductively coupled plasma (ICP)
etching, with a Cl$_2$/Ar mixture. Finally, the sacrificial layer
and remaining SiN mask were removed with a $>10$~s, 49 $\%$ HF dip.
In many devices, the SiN$_x$ mask layer was omitted and direct
transfer from the electron-beam mask to the GaAs was performed using
the same ICP etch.  No significant change in device quality or
performance was observed in going from a straight electron-beam mask
to a SiN$_x$ mask.

For device interrogation within the cryostat, it was often most
convenient to isolate the waveguide devices to a mesa that was
$\approx$15 $\mu$m above the rest of the sample's surface.  This was
done through contact photolithography and a
H$_2$O$_2$:H$_3$PO$_4$:CH$_3$OH (10:1:1 by volume) solution at
50~$^\circ$C for 30~min.

The material containing dots with s-shell emission near 940~nm was
annealed in a rapid thermal annealer at $\approx830^\circ$C for 30~s
prior to fabrication. This was done to blue-shift the QD s-shell
emission, which originally occurred at $\approx1100$ nm. In
addition, a first set of measurements revealed that fabricated
waveguides were insufficiently narrow to achieve phase-matching to
the FTW in a wavelength-band matching the QD emission. To rectify
this, a digital GaAs wet etch technique\cite{ref:Hennessy2} was
employed to reduce the waveguide dimensions appropriately. The
process consisted of alternately oxidizing the GaAs in H$_2$O$_2$
and removing the formed oxide layer with a 1 molar solution of
citric acid (C$_6$H$_8$O$_7$).

\subsection*{Transmission spectrum measurement}
\begin{figure}[h]
\centerline{\includegraphics[width=15cm]{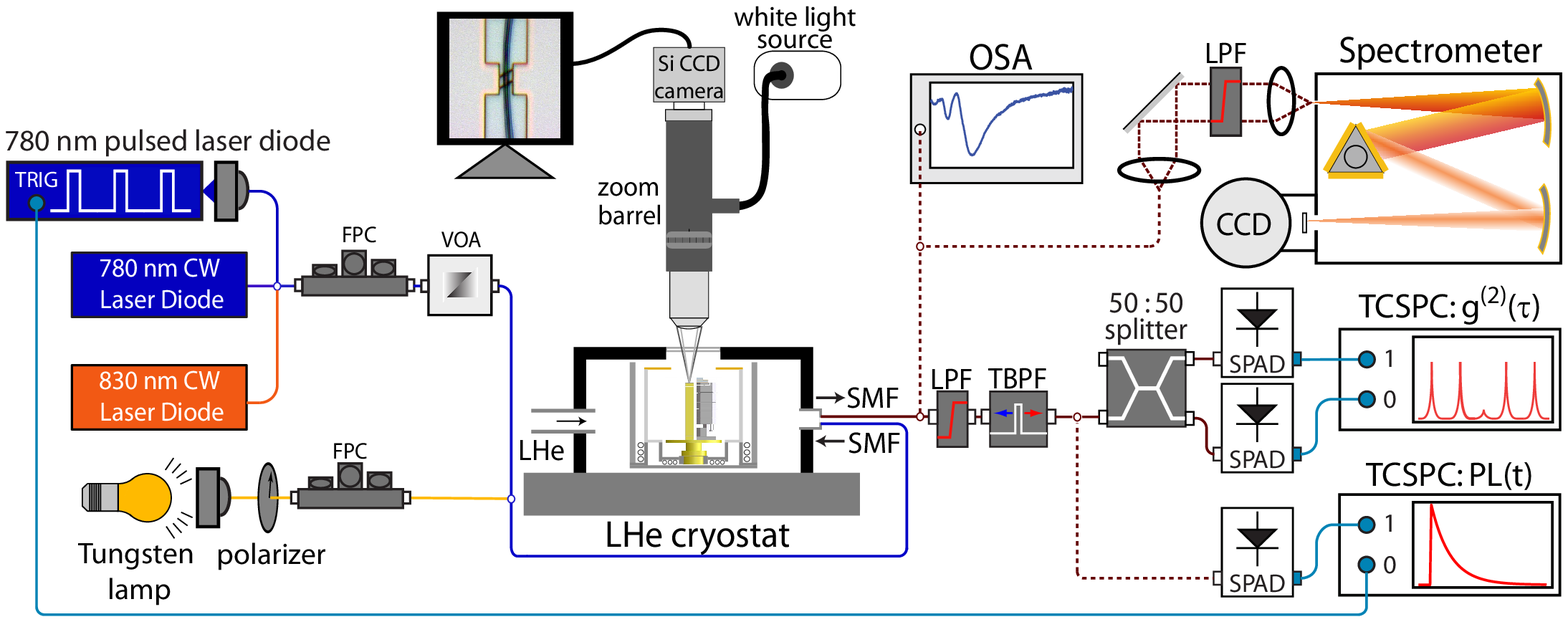}}
\caption{Experimental setup for cryogenic transmission and
photoluminescence measurements. VOA: variable optical attenuator;
SMF: single mode optical fiber OSA: optical spectrum analyzer; LPF:
long wavelength pass filter; TBPF: tunable bandpass filter; FPC:
fiber polarization controller; SPAD: single photon avalanche
detector; TCSPC: time correlated single photon counter; CCD:
charge-coupled device; LHe: liquid helium; TRIG: trigger. }
\label{Fig:exp_setup}
\end{figure}

As illustrated in Fig.~1 in the main text, a FTW of $\approx 1~\mum$
diameter was brought into contact with individual waveguides,
forming directional coupler structures that are interrogated using
the experimental setup depicted in Fig.~\ref{Fig:exp_setup}. Light
from a quartz-tungsten-halogen lamp was coupled into a single-mode
optical fiber and passed through an in-line polarizer and
polarization controller, and then launched into the FTW input. An
optical spectrum analyzer (OSA) was used to obtain transmission
spectra, from 1100 nm to 1600 nm, for the formed directional
couplers. To obtain spectra for the two main polarizations, the
polarization controller was used to minimize transmission at some
wavelength range after the WGs were first put in contact. This range
was assumed to be that for which power transfer from the suspended
WG to the fiber was maximized, for one of the two main
polarizations. A spectrum was recorded, the fiber lifted and a
background spectrum was then taken. The fiber was next brought back
in contact with the GaAs guide, and a second spectrum was taken to
verify that the transmission minima, and thus the polarization, was
unchanged. The polarization controller was next used to maximize the
transmission at the minimum wavelength range, and a second
transmission spectrum was taken, assumed to be for the second main
polarization. Finally, the fiber was lifted and a second background
spectrum was recorded.

The minimum transmission wavelengths plotted on Fig.~2(b) in the
main text correspond to measured global transmission minima.
Detection noise leads to uncertainties in the determination of an
actual minimum transmission wavelength. This is represented by error
bars in the figure, which correspond to intervals over which the
transmitted power fluctuates below 1.05 times the minimum
transmission.

To determine the waveguide widths, top view scanning electron
microscope images of the devices were taken, and an edge detection
technique was employed to determine sidewall profiles along the
waveguide length. For each waveguide, averages and standard
deviations were obtained for the locations of the two waveguide
sidewalls, which were then used to calculate waveguide widths and
corresponding deviations due to sidewall roughness. The standard
deviations are plotted as error bars in the graph of minimum
transmission wavelength versus waveguide width on Fig.~2(b) in the
main text.
\subsection*{Determination of collection efficiency}
Collected photon rates were determined from PL spectra, after
calibration of Si CCD count rates against the known optical power of
a (fiber-coupled) continuous wave laser source, tuned to the QD
emission wavelength. Integrated counts in the spectrum of a 100~fW
laser signal led to a conversion factor of $130$ photons/CCD count.
We point out that this conversion factor, which includes the
spectrometer in-coupling efficiency, grating efficiency, and CCD
detection efficiency and gain, falls within the expected range,
considering manufacturer-provided specifications. The Si CCD quantum
efficiency is $<10~\%$ at 960 nm, the CCD gain was set to 3
photoelectrons/CCD count, and the grating efficiency is around
50~\%.

When measuring PL, a long-wavelength pass filter at 850~nm was
introduced before the spectrometer slit, with a nominal transmission
of $80~\%$. With this, the collected photon rate into the single
mode fiber was calculated as $R_{ph.}=R{_{det.}}\cdot130/0.8$. To
determine the emitted photon rate, the QDs were pumped into
saturation and assumed to have $100~\%$ radiative efficiency. Under
these conditions, the emission rate is that of the pump source,
50~MHz.  The collection efficiency into the single mode fiber,
$\eta$, is then given as $\eta=R_{ph.}/50{\times}10^6$.

Considering an ideal single exciton QD line, the collected photon
rate $R_{ph}$ is given by the product of the collection efficiency
$\eta$, the pulse repetition rate $R_{rep.}=50$~MHz and the average
exciton occupancy per pulse: $R_{ph} = \eta\cdot
R_{rep}/(1+P_{in}/P_{sat})$, where $P_{in}$ is the average pump
power and $P_{sat}$ is the pump saturation power. This model fits
the data of Fig.~3(c) reasonably well for $P_{in}<100$~nW, with
$P_{sat}=133.2$~nW$\pm21.7$~nW (the uncertainty is a 95~\% fit
confidence interval, associated with errors in the determination of
the emission rates of the isolated excitonic line). For
$P_{in}>100$~nW, however, it overestimates the output PL rate, and
for $P_{in}>580$~nW, a marked decline in PL intensity is observed,
suggesting that the emission rate starts to decrease before QD
saturation is achieved.
\subsection*{$g^{(2)}(\tau)$ measurements}
A fiber-based Hanbury-Brown and Twiss configuration was used to
obtain the second-order correlation curve shown in Fig.~3(e) in the
main text. The excitation was pulsed and below saturation
($P_{in}\approx 75$~nW). The setup consisted of two Si single photon
avalanche detectors (SPADs) connected to the output ports of a 3dB
optical fiber splitter, and a time-correlated single photon counting
instrument, which performed histogramming of the photon arrival time
differences. The SPAD detection efficiency was determined to be
$26~\%\pm4~\%$ and the splitter transmission $37~\%\pm1~\%$ per arm.
The PL signal from the FTW was filtered with a fiber coupled, 900~nm
long-wavelength pass filter, followed by a fiber coupled, tunable,
thin-film filter with a 1~nm bandwidth, with a $55~\%\pm4~\%$
transmission. The total transmission through the two filters was
$26~\%\pm2~\%$.

To obtain the $g^{(2)}(0)$ value quoted in the text, we integrated
the peaks in Fig.~3(e) in the main text over increasing intervals up
to half the repetition period, and calculated averages. The mean and
standard deviation values for the zero time peak (which corresponds
to $g^{(2)}(0)$) were then normalized by an average of all other
integrated (averaged) peak values.

To estimate the percentage of background emission in the analyzed
signal, we derived an expression for $g^{(2)}(\tau)$ for the case of
detection of QD emission together with perfectly uncorrelated
background light. The second order correlation function for the
total detected light is
\begin{equation}
g^{(2)}(\tau) = \frac{\langle I(t)I(t+\tau)\rangle}{\bar{I}^2},
\end{equation}
where $I(t)=I_{QD}(t)+I_{bg}(t)$ is the total detected intensity,
$I_{QD}$ and $I_{bg}$ are the quantum dot emission and background
intensities respectively, and $\bar{I} = \langle I(t) \rangle$.
Assuming \begin{equation}\langle I_{QD}(t)I_{bg}(t+\tau)
\rangle=\langle I_{bg}(t+\tau)I_{QD}(t) \rangle =
\bar{I}_{QD}\bar{I}_{bg},\end{equation} we get
\begin{equation}
g^{(2)}(\tau) = \frac{\left(
\bar{I}^2-\bar{I}_{QD}^2\right)+\bar{I}_{QD}^2\cdot
g^{(2)}_{QD}(\tau)}{\bar{I}^2},
\end{equation}
with $g^{(2)}_{QD}(\tau)=\langle I_{QD}(t)I_{QD}(t+\tau)\rangle$.
From this expression, the zero-time second-order correlation for the
QD alone is
\begin{equation}
g^{(2)}_{QD}(0)=1+\left(\frac{1+R}{R}\right)^2\cdot[g^{(2)}(0)-1],
\end{equation}
where $g^{(2)}(0)$ is the measured quantity, and $R=I_{QD}/I_{bg}$
is the ratio of QD light intensity $I_{QD}$ over background light
intensity $I_{bg}$ at the detector (i.e., R is the
signal-to-background ratio). Since $I_{QD}+I_{bg}=I$, we get
\begin{equation}
\frac{I_{bg}}{I_{total}} = \frac{1}{1+R}.\label{bgratio}
\end{equation}
Assuming perfectly antibunched single QD emission, $R$ was obtained
by setting $g^{(2)}_{QD}(0)=0$, and the background ratio over the
total was computed from~\ref{bgratio}.

Assuming that the detected signal is composed of perfectly
antibunched light from the single QD emission peak at 963~nm and
uncorrelated background photons, we determine that $16~\%$ of the
total detected photons would correspond to background emission. In
this case, since the area under the sharp excitonic line in the
spectrum of the main text's Fig.~3(e) inset is $\approx51~\%$ of the
total, the broad features would have to be produced in part by the
same QD. A more likely situation is that the detected signal also
includes emission from additional (broadened) QDs.
\section*{Photoluminescence excitation spectroscopy}
In addition to photoluminescence spectroscopy, our fiber-based
technique can be used for photoluminescence excitation (PLE)
spectroscopy of single quantum dots, even in samples with a high
density of emitters. In such a measurement, the QD is excited
quasi-resonantly in the p-shell, by a continuous wave, tunable
external cavity diode laser (ECDL), and the s-shell PL is recorded
as the excitation wavelength is scanned. This type of measurement
has been used in the past to reveal carrier scattering mechanisms
leading to s-shell photon emission~\cite{ref:Toda2,ref:Warming}.

\begin{figure}[t]
\centerline{\includegraphics[width=12cm,trim= 0 0 0
0]{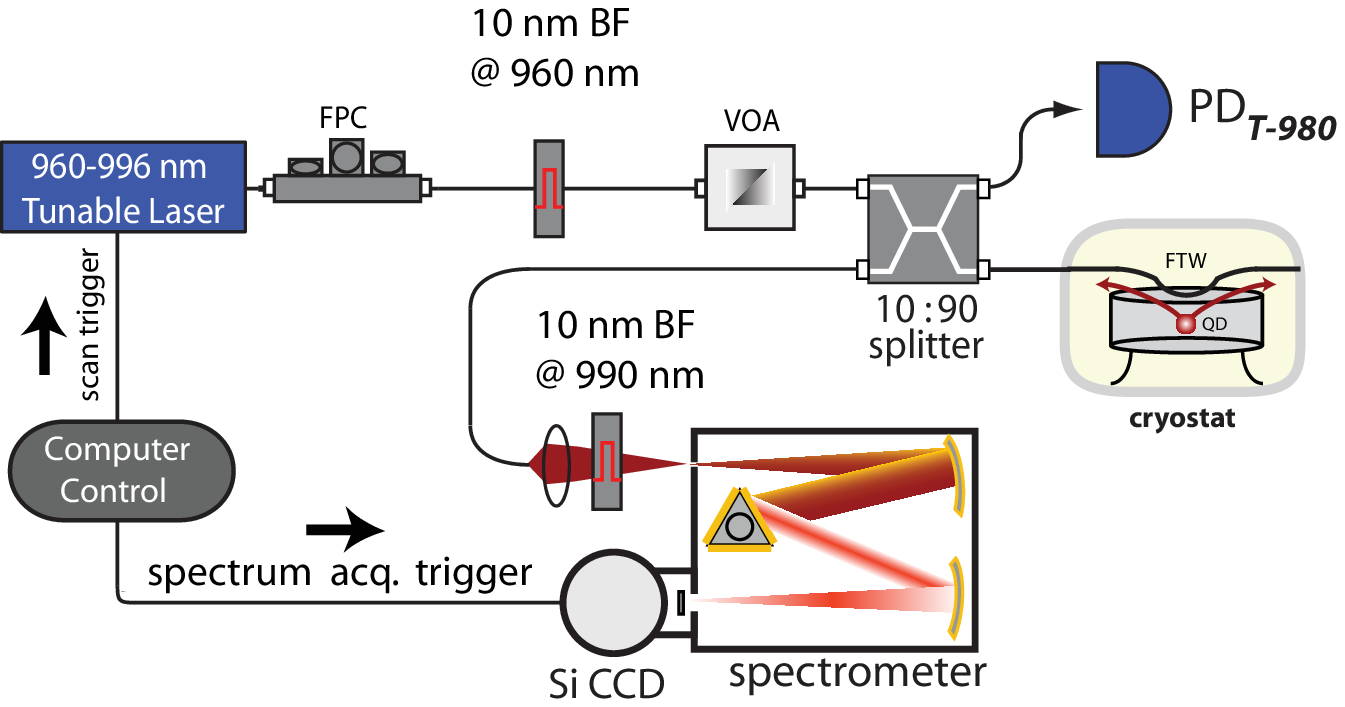}} \caption{Photoluminescence excitation
spectroscopy setup} \label{Fig_PLE_setup}
\end{figure}

\begin{figure}[!]
\centerline{\includegraphics[width=\linewidth,trim= 0 0 0
0]{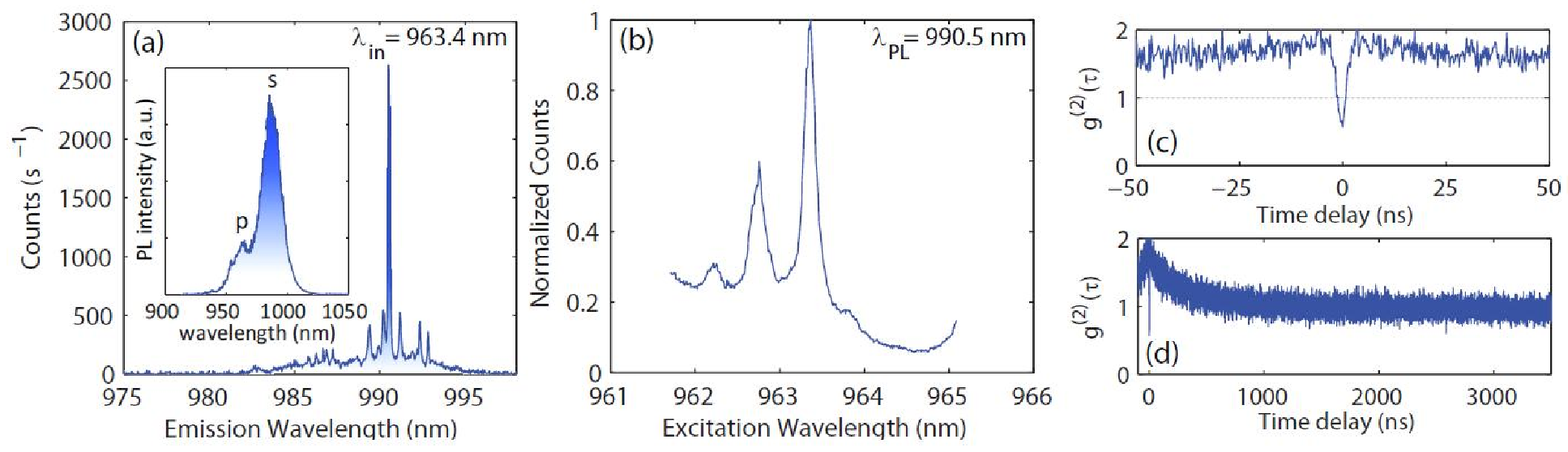}} \caption{(a) Fiber-collected PL spectrum for CW
quasi-resonant excitation at $\lambda=963.4$~nm. Inset: ensemble QD
non-resonant PL spectrum, obtained via free-space collection. (b)
  Evolution of  PL intensity for the single dot at $\lambda = 990.5$ nm, for varying excitation wavelengths. (c) CW second-order photon correlation
trace for the PL peak at $\lambda = 990.5$ nm in (a). (d) same as
(c), with longer time span. } \label{Fig_PLE}
\end{figure}

We next describe a single QD PLE measurement performed on a high QD
density sample, with s-shell emission near 1000 nm, as shown in the
inset of Fig.~\ref{Fig_PLE}~(a). In our experimental setup,
Fig.~\ref{Fig_PLE_setup}, the QD was excited with continuous wave
light from a 963 nm to 995 nm ECDL via the FTW. The emitted s-shell
PL, collected back into the input fiber (i.e., flowing in the
direction opposite to that of the excitation light), was diverted
towards a grating spectrometer through a 10:90 directional fiber
coupler. This was done to reduce the intensity of excitation light
at the spectrometer entrance, which would otherwise have overwhelmed
the PL signal. The laser wavelength was set to be scanned over a
portion of the QD ensemble p-shell range, with a rate of 10 pm/s.
During the scan, photoluminescence spectra with a 1 s integration
time were continuously recorded.
Figure~\ref{Fig_PLE}(a) shows fiber collected s-shell emission for
an excitation wavelength $\lambda_{\rm{in}}=963.4$~nm, for which a
bright, single QD line at $\lambda=990.5$~nm becomes visible. The
evolution of this line's intensity for $\lambda_{\rm{in}}$ varying
between 961.5 nm and 965 nm is shown Fig.~\ref{Fig_PLE}(b), with a
spectral resolution better than 30 pm. The excitation spectrum
displays sharp peaks related to the carrier dynamics leading to
s-shell emission~\cite{ref:Toda2,ref:Warming}.

To confirm that the selected PL line is indeed produced by single
dot, photon correlation measurements are performed with a
Hanbury-Brown and Twiss setup, for $\lambda_{\rm{in}}=963.4$~nm. A 1
nm, fiber-based tunable bandpass filter was used to filter the PL
signal. The antibunching dip at zero delay ($g^{(2)}(0) \approx
0.5$) shown in Fig.~\ref{Fig_PLE}(c) confirms that the p-shell
excitation through the coupler can indeed select a single QD from
within the dense ensemble. The nonzero antibunching is most likely
due to the limited measurement resolution (256 ps) and insufficient
bandpass filtering, especially considering the high QD density of
the sample. Bunching peaks surrounding the antibunching dip are also
observed (Fig.~\ref{Fig_PLE}(d)), with a decay time of a few
hundreds of nanoseconds, and are possibly related to the formation
of charged QD states~\cite{ref:Santori4}.
\bibliographystyle{apsrev}

\end{document}